\documentclass[twocolumn,showpacs,preprintnumbers,amsmath,amssymb,prl,superscriptaddress]{revtex4-1}

\usepackage{graphicx}
\usepackage{bm}

\begin{document}

\title{Entanglement witnessing in superconducting beamsplitters}
\author{H. Soller}
\affiliation{Institut f\"ur Theoretische Physik, Ruprecht-Karls-Universit\"at Heidelberg,\\
 Philosophenweg 19, D-69120 Heidelberg, Germany}
\author{L. Hofstetter}
\affiliation{Department of Physics, University of Basel, Klingelbergstrasse 82, CH-4056 Basel, Switzerland}
\author{D. Reeb}
\affiliation{Department of Mathematics, Technische Universit\"at M\"unchen, 85748 Garching, Germany}
\date{\today}

\begin{abstract}

We analyse a large class of superconducting beamsplitters for which the Bell parameter (CHSH violation) is a simple function of the spin detector efficiency. For these superconducting beamsplitters all necessary information to compute the Bell parameter can be obtained in Y-junction setups for the beamsplitter. Using the Bell parameter as an entanglement witness, we propose an experiment which allows to verify the presence of entanglement in Cooper pair splitters.

\end{abstract}


\maketitle

\begin{figure}
\centering
\includegraphics[width=7cm]{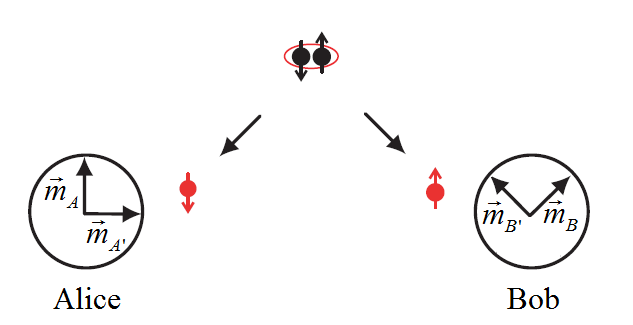}
\caption{A schematic for a Bell measurement on the electrons of a Cooper pair. First, the two electrons with opposite (entangled) spin are separated to the observers Alice and Bob. Afterwards projection measurements along the axes $\mathbf{m}_{A,A'}$ and $\mathbf{m}_{B,B'}$ are performed.}
\label{fig3}
\end{figure}

Cooper pairs in conventional superconductors form spin singlet states, which makes them in principle an ideal on-chip source of entangled electron pairs. Everything that has to be done is to coherently transfer the electrons from the superconductor (where all electrons are in the same condensate) to two spatially separated places in order to obtain a maximally entangled Einstein-Podolsky-Rosen (EPR) pair of electrons \cite{springerlink:10.1007/BF01491987}. The Cooper pairs can be extracted from the superconductor via tunneling but splitting of the pair into separate electrodes has to be enforced, which requires a high degree of control of the device. The most promising experimental realisations of superconducting beamsplitters have focussed on controlling the extraction of electrons by Coulomb interaction in tunable double quantum dots \cite{PhysRevLett.104.026801,19829377,PhysRevLett.91.267003}. In very recent experiments\cite{PhysRevLett.107.136801,2009arXiv0910.5558W,das2012high} it has been clearly demonstrated that Cooper pairs can be split into separate electrons in two normal metals. However, so far no experiment has demonstrated that these objects are indeed entangled pairs.\\
Most proposals have employed one of two ways to establish the presence of entanglement: one can use the fermionic analogue of the Hong-Ou-Mandel dip interferometry\cite{PhysRevLett.84.1035,PhysRevB.61.R16303,PhysRevB.74.165326} or four-terminal devices to realize a Bell measurement device \cite{PhysRevLett.94.210601,PhysRevB.66.161320,PhysRevB.83.125304} or an entanglement witness \cite{PhysRevB.75.241305}. While the former proposal relies on a direct consequence of particle indistinguishability in quantum mechanics the latter allows for a direct measurement of quantifiable entanglement, which is what we want to pursue here. All proposals rely on measuring higher order cumulants of the current flow in the beamsplitter with the exception of Ref. \cite{PhysRevLett.94.210601}, which discusses a time-resolved measurement of electron spins arriving at the spatially separated observers usually referred to as Alice and Bob. Experimentally, however, such measurements have proven to be very complicated \cite{Henny09041999}. Additionally, all proposals so far require a violation of Bell's inequality which is actually not needed in order to verify the presence of entanglement \cite{PhysRevA.40.4277}.\\
In this paper we approach entanglement witnessing in superconducting beamsplitters in two ways: on the one hand we will show that for a large class of four-terminal setups the Bell parameter will only depend on the efficiency of the detector. This analysis will allow us to go over to a simpler Y-junction geometry. On the other hand we will show that turning the Bell parameter into a weaker entanglement witness will considerably faciliate the observation of entanglement. Both findings together will put us in the position to show that detection of entanglement in superconducting beamsplitters is possible with present day technique if integrated in a single experiment.\\
We start with the simplest setting as in \cite{PhysRevLett.94.210601}. Both Alice and Bob can perform spin measurements in two directions $\mathbf{m}_A, \mathbf{m}'_A$ and $\mathbf{m}_B, \mathbf{m}'_B$, respectively, see Fig. \ref{fig3}. 
\\
As in optical experiments we define the Bell parameter (from the CHSH inequality) to be 
\begin{eqnarray}
\epsilon &=& |E(\mathbf{m}_A,\mathbf{m}_B) + E(\mathbf{m}'_A,\mathbf{m}_B) \nonumber\\
&& + E(\mathbf{m}_A,\mathbf{m}'_B) - E(\mathbf{m}'_A,\mathbf{m}'_B)|, \label{bp}
\end{eqnarray}
\begin{figure*}
\centering
\includegraphics[width=14cm]{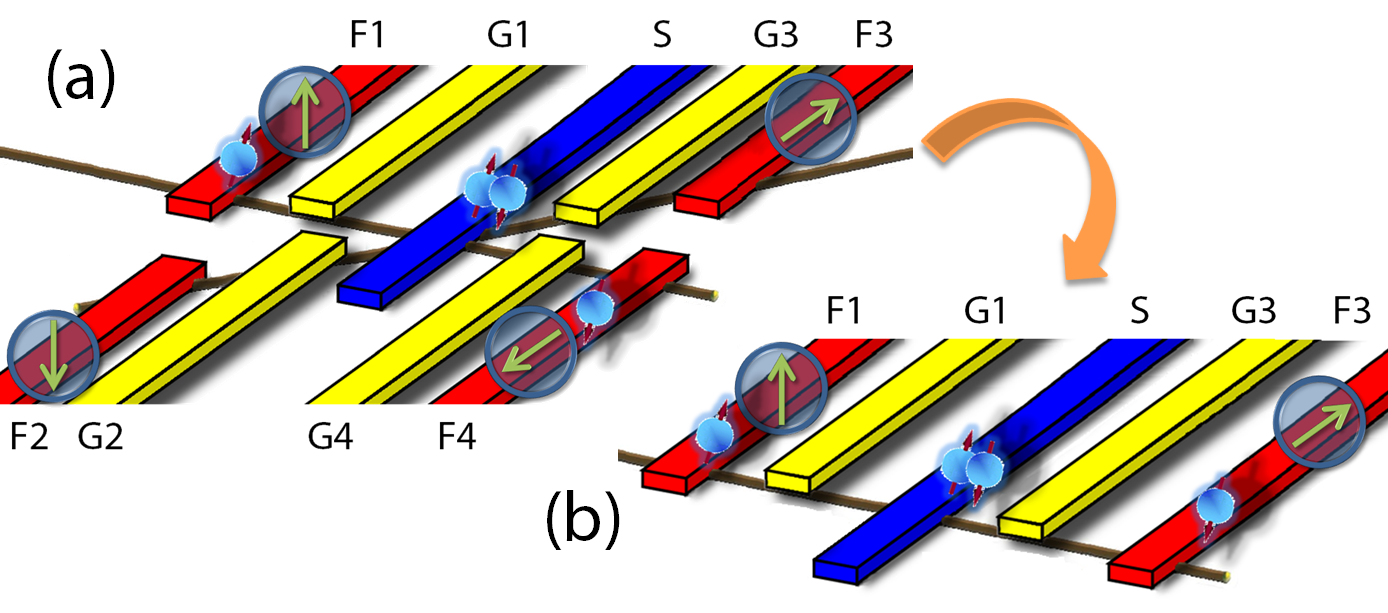}
\caption{Sketch of the experimental setups. (a): a central superconducting finger $S$ (blue) is contacted to two InAs nanowires (brown). These in turn are contacted by four ferromagnetic electrodes (red), the direction of magnetization of which are indicated by arrows here for the configuration of $\mathbf{m}_A$ and $\mathbf{m}_B$ as in Fig. \ref{fig3}. Alice and Bob are represented by the spin detectors consisting of electrodes $F1$, $F2$ and $F3$, $F4$ respectively. The emerging quantum dots between the superconducting finger and the four ferromagnetic electrodes are tunable by top gates $G1$-$G4$. (b): using our analysis the setup can be considerably simplified to a Y-junction geometry.}
\label{fig1}
\end{figure*}\\
where the correlator is given by $E(\mathbf{m},\mathbf{m}') = P_{\mathbf{m}\mathbf{m}',++} + P_{\mathbf{m}\mathbf{m}',--} - P_{\mathbf{m}\mathbf{m}',+-} - P_{\mathbf{m}\mathbf{m}',-+}$. $P_{\mathbf{m}\mathbf{m}',\sigma\sigma'}$ denotes the probability of observing an electron pair in detectors with directions $\mathbf{m},\mathbf{m}'$ with spin directions $\sigma=\pm$ and $\sigma'=\pm$. In our case the detectors for electrons will be the four ferromagnetic terminals ($F1$ - $F4$) and the source will be the superconductor ($S$), see Fig. \ref{fig1}(a). The density of states of a ferromagnet is spin dependent and can be written as $\rho_{0\sigma} = \rho_0 (1+ \sigma P)$, where $P, \; |P| \leq 1$ is the polarisation.  Therefore the detection of an electron in a ferromagnet is also an (imperfect) spin measurement. We write the magnetisation direction $\mathbf{g}_i, \; i = 1,2,3,4$ as $\mathbf{g}_i = \mathbf{m}_i P_i$ with a unit vector $\mathbf{m}_i$ that indicates the direction. We assume for notational simplicity that all four ferromagnets attached have the same polarisation and the same coupling to the superconductor. Additionally, in order to have spin detectors, the magnetisation direction of ($F1$, $F2$) and ($F3$, $F4$) is assumed to be pairwise antiparallel as also indicated in Fig. \ref{fig1}(a). We therefore write $\mathbf{g}_1 = - \mathbf{g}_2 = \mathbf{g}_A, \; \mathbf{g}_3 = - \mathbf{g}_4 = \mathbf{g}_B$ referring to Alice and Bob. The case of non-equal polarisations in a special case has been treated in \cite{0295-5075-81-4-40002}. It does not lead to a qualitatively different behavior.\\
In a first approximation for small bias $V \ll \Delta$, $\Delta$ being the superconductor gap, let us consider processes where a Cooper pair comes from the superconductor, is split and transferred further to the separate leads without a specific energy dependence of SC DOS. This transfer is enforced by applying the bias voltage $V$ between the leads $F1$ - $F4$ and $S$. The leads are all kept at the same electro-chemical potential. Then we can write the conductance $G= dI/dV$ for small bias $V$ and for transfer from $S$ to $Fi$ and $Fj$ as (using units such that $e = \hbar = k_B = 1$)
\begin{eqnarray}
G_{ij} = f(V, \delta r, \Sigma) (1- \mathbf{g}_i \mathbf{g}_j), \label{gij}
\end{eqnarray}
where $f(V, \delta r, \Sigma)$ is a function of the applied bias $V$, the width $\delta r$ of the superconductor (superconducting finger in Fig. \ref{fig1}) and a possible on-site interaction on the quantum dot described by a self-energy $\Sigma$. For the case of simple tunnel contacts between the superconductor and the ferromagnets $f(V, \delta r, \Sigma)$ is just a constant $f(V, \delta r, \Sigma) = 16\Gamma^2/(1+4\Gamma)^4$, where $\Gamma$ is the (dimensionless) transparency of the contact between the superconductor and the ferromagnets \cite{PhysRevLett.94.210601}. This result remains valid even in the case of finite temperature. It also gives the probabilities $P_{ij}$ for simultaneous detection of an electron at $Fi$ and $Fj$ by \cite{PhysRevLett.94.210601}
\begin{eqnarray}
P_{ij} = G_{ij} / \sum_{\{k,l\}} G_{kl}, \label{pab}
\end{eqnarray}
where the sum over $\{k,l\}$ is over all pairs of detectors considered. We will take into account only detection events that involve both Alice and Bob in order to obtain the correlator in Eq. (\ref{bp}), so that $i = 1,2$ and $j=3,4$ and consequently disregard events in which both electrons go to the same ferromagnet or both go to Alice. Since the ferromagnets' polarisations are antiparallel for Alice and Bob we associate the detection of an electron in $i=1, \; j=3$ with a measurement $+$ and $i=2, \; j=4$ with a measurement $-$. As we only count events that involve both Alice and Bob we can immediately calculate the probabilities $P_{\mathbf{m}\mathbf{m}',\pm\pm}$ since if in such an event an electron does not arrive at e.g. $i=1$ it has to go via $i=2$ due to our choice of events. Performing the sum in Eq. (\ref{pab}) using our expression from Eq. (\ref{gij}) leads to a simple expression $E(\mathbf{m}_A,\mathbf{m}_B) = - \mathbf{g}_A \mathbf{g}_B$ for arbitrary $f(V, \delta r, \Sigma)$. Using this result the Bell parameter is given by
\begin{eqnarray}
\epsilon &=& P^2 \epsilon_0, \label{bell} \\
\epsilon_0 &=& |\mathbf{m}_A \mathbf{m}_B + \mathbf{m}'_A \mathbf{m}_B + \mathbf{m}_A \mathbf{m}'_B - \mathbf{m}'_A \mathbf{m}'_B|. \label{e0}
\end{eqnarray}
The maximum value for $\epsilon_0$ for an appropriate choice of angles (as shown in Fig. \ref{fig3}) between the different measurement directions is $2 \sqrt{2}$, the so-called Tsirelson's bound \cite{springerlink:10.1007/BF00417500}. A violation of Bell's inequality is reached for $\epsilon > 2$ which requires the polarisation $P$ to be at least $84\%$.\\
The calculation above does not only apply to a tunnel contact but to all systems whose conductance has the form of Eq. (\ref{gij}). We now want to show that this is the case for a large class of systems by considering several possible generalizations of the setup described above.\\
In \cite{0295-5075-81-4-40002} diffusive charge transfer instead of ballistic transport through a tunnel contact as discussed above was considered, however, Eq. (\ref{gij}) is recovered with a modified, but constant, $f(V, \delta r, \Sigma)$. Consequently the simple form of the Bell parameter in Eq. (\ref{bell}) is recovered independently from the charge transfer.\\
In the next step we consider a resonant level (a single energy level at $\Delta_0$ without interactions) in between the ferromagnets and the superconductor in order to study effects of energy dependent tunneling and finite voltage bias. In \cite{soller1} a resonant level coupled to four ferromagnets was considered and the conductance for the events involving both Alice and Bob is given by
\begin{eqnarray}
&& G_{R,ij}(V) = [T_{R,ij}(V) + T_{R,ij}(-V)] (1- \mathbf{g}_i \mathbf{g}_j),\label{grij}
\end{eqnarray}
where
\begin{eqnarray}
&& T_{R,ij}(\omega) = \left\{16 \Gamma_F^2 \Gamma_S^2\right\} / \left\{(\omega - \Delta_{\sigma})^2 (\omega - \Delta_{-\sigma})^2 \right. \nonumber \\
&& + (4 \Gamma_F^2 + \Gamma_S^2)^2 + 4 \Gamma_F^2 [(\omega - \Delta_{\sigma})^2 + (\omega - \Delta_{-\sigma})^2] \nonumber\\
&& \left.  + 2 \Gamma_S^2 (\omega - \Delta_\sigma)(\omega - \Delta_{-\sigma}) \right\}, \nonumber
\end{eqnarray}
where $\Gamma_F, \; \Gamma_S$ are the tunneling rates from the quantum dot to the ferromagnet and the superconductor respectively. $\Gamma_S = \Gamma_{S0} \Delta /\sqrt{\Delta^2 - \omega^2}$, involves the superconductor gap $\Delta$. $\Delta_\sigma = \Delta_0 - \sigma h$ is the position of the resonant level $\Delta_0$ that may be split by an applied magnetic field / exchange field $h$. The conductance in Eq. (\ref{grij}) is again of the form in Eq. (\ref{gij}).\\
Up to now we only considered a single resonant level not including any dependence on a finite length $\delta r$ and a non-interacting system. However, treating several quantum dots only leads to more elaborate expressions for the denominator of $T_{R,ij}$ in Eq. (\ref{grij}) but no difference in the dependence on the magnetisation directions is involved \cite{soller1}. Therefore, in order to proceed analytically and for reasons of clarity we focus from now on systems with a single dot at zero bias.\\
First we consider the finite length of the nanotubes/nanowires typically used in the experiments to form the quantum dots (for semiconductors see \cite{2007NatPh...3..455Y,0295-5075-54-2-255}, for carbon nanotubes see \cite{PhysRevB.84.115448}): the tunneling rate between the superconductor and the quantum dot acquires a dependence on the width $\delta r$ of the superconducting finger of the kind $[\sin (k_F \delta r) /(k_F \delta r)]^2 \exp[-2 \delta r /(\pi \xi)]$, where $k_F$ refers to the Fermi velocity in the superconductor and $\xi$ is the superconductor coherence length. Inlcuding these rescaled tunnel rates in Eq. (\ref{grij}) the conductance remains of the form of Eq. (\ref{gij}). We note that not coupling the superconductor directly to the quantum dot but using a topological insulator which acquires a superconducting gap via the proximity effect of a superconducting slab deposited on top leads to a similar dependence on the width $l$ of the superconducting slab in the conductance expression \cite{PhysRevB.82.081303}, leaving it to be of the form in Eq. (\ref{gij}).\\
The last example includes spin-independent interactions. We follow \cite{PhysRevB.63.094515} and generalize the result of one normal terminal to the case of four ferromagnets attached to a superconductor via a quantum dot. In this case zero-bias conductance is given by
\begin{eqnarray}
G_{I,ij} = \frac{16 \tilde{\Gamma}_s^2 \Gamma_F^2 (1- \mathbf{g}_i \mathbf{g}_j)}{(\tilde{\Delta}_0^2 + (4\Gamma_F)^2 + \tilde{\Gamma}_s^2)^2}, \label{giij}
\end{eqnarray}
where $\tilde{\Gamma}_S = \Gamma_{S0} - \Sigma_{12}(0)$ and $\tilde{\Delta}_0 = \Delta_0 - \Sigma_{11}(0)$. $\Sigma_{11}(0)$ and $\Sigma_{12}(0)$ represent the 11- and 12-component of the Nambu self-energy at zero energy \cite{PhysRevB.63.094515}. We do not want to specify the type of interaction any further but interactions with a local phonon would be a typical example. The conductance is again of the form of Eq. (\ref{gij}). We will discuss a typical example of spin-dependent interactions later.\\
Therefore we have shown that Eq. (\ref{bell}) is valid for a large class of systems independent of the type of charge transfer, temperature, magnetic fields, spin-independent interaction or finite voltage. This is the first important result of this paper and can be explained by the fact that the observable probabilities are normalized in Eq. (\ref{pab}) with respect to the events where a Cooper pair is split. Therefore, the rate of such processes, which is of course affected by the above mentioned effects, does not enter Eq. (\ref{pab}).\\
We conclude from this analysis that the nonlocal conductances have the form of Eq. (\ref{gij}). As they only depend on the pairwise alignment of the polarisation we can obtain all required nonlocal conductances also from just two ferromagnets attached to the superconductor \cite{PhysRevLett.94.210601}. Consequently we can also work with a typical Y-junction geometry like the splitter indicated in Fig. \ref{fig1}(b).\\
However, several remarks concerning this result need to be made. The first being that Eq. (\ref{gij}) assumes that the correlated electron pairs do not suffer any corruption along the way. This is a justified assumption as far as transport in the superconductor and via the quantum dots is concerned since in the superconductor the Cooper pair remains coherent and the double quantum dot structure has dimensions of $\approx 100$nm \cite{19829377} so that spin relaxation shouldn't be relevant \cite{PhysRevLett.99.036801}. Nonetheless, manifold processes may happen at the interface between the ferromagnet and the quantum dot leading to spin-active scattering \cite{soller4}. If such processes would only lead to a reduction of the nonlocal processes (and consequently the nonlocal conductance in Eq. (\ref{gij})) they would not matter. However, their presence leads to additional contributions to the nonlocal conductance which are due to the splitting of spin-flipped Cooper pairs \cite{Soller2012,PhysRevLett.110.047002}. In the presence of such processes our simple analysis from above does not apply.\\
Additionally, polarisations much higher than $P\approx40\%$ are hard to reach with present day materials \cite{coey}. On the other hand, there is a difference between the verification of entanglement and the actual violation of a Bell inequality, which implies a violation of 'local reality' \cite{PhysRevA.40.4277}. Everything we need to know about the quantum state in question is whether it is not separable, meaning that it cannot be written as a convex combination of tensor product states. Generically separability implies stronger inequalities than local reality \cite{PhysRevLett.94.010402,Uffink20081205}. For any locally realistic theory the Bell parameter in Eq. (\ref{bp}) has to be smaller than 2. Now let us assume that Alice and Bob's two measurement directions for spin are orthogonal, meaning e.g. that $\mathbf{m}_A$ points in $x$-direction whereas $\mathbf{m}'_A$ points in $y$-direction and similar for Bob. In this case $\epsilon_0$ in Eq. (\ref{e0}) can still reach the maximal value \cite{springerlink:10.1007/BF00417500} $2\sqrt{2}$ (e.g. for $\mathbf{m}_A = \hat{\mathbf{e}}_x, \; \mathbf{m}'_A = \hat{\mathbf{e}}_y, \; \mathbf{m}_B = 1/\sqrt{2} [\hat{\mathbf{e}}_y + \hat{\mathbf{e}}_x]$ and $\mathbf{m}'_B = 1/\sqrt{2} [\hat{\mathbf{e}}_y - \hat{\mathbf{e}}_x]$, see Fig. \ref{fig3}). However, the maximal value for the Bell parameter in Eq. (\ref{bp}) for a separable quantum state is now only $\sqrt{2}$ \cite{PhysRevLett.94.010402}. Therefore witnessing an entangled quantum state via Eq. (\ref{bell}) only requires $P^2 2 \sqrt{2} \geq \sqrt{2} \Rightarrow P \approx 70\%$. This means the following: if we restrict ourselves to merely witnessing entanglement and if we trust our device to measure a specific set of spin directions without the presence of spin-active scattering, then the requirement on the polarisation strength is much less severe. The absence of spin-active scattering is a strong requirement since while it is typically not present in the normal-superconductor beamsplitters \cite{PhysRevLett.107.136801,2009arXiv0910.5558W,das2012high} it is a general feature of superconductor-ferromagnet heterostructures \cite{PhysRevB.81.094508}. We will discuss below one specific case where it can be neglected.\\
In the last part we want to describe an actual experiment which is feasible and able to witness entanglement in a superconducting beamsplitter. The authors of \cite{PhysRevLett.94.210601} were aiming at the measurement of the probabilities in (\ref{pab}) via the detection of single events in the spin detectors and averaging over them. However, these time-resolved coincidence measurements are not necessary since the probabilities can also be obtained by measuring the non-local conductances and calculating the probabilities from them, as described above. Experimental access to the non-local conductance in superconducting beamsplitters has been demonstrated \cite{PhysRevLett.107.136801} via measuring the total conductance between the superconductor and one of the leads (e.g. $F1$ in Fig. \ref{fig1}(b)). The non-local contribution between two leads (e.g. $F1$ and $F3$) can now be obtained by varying the bias of the second lead ($F3$ in Fig. \ref{fig1}(b)) and keeping only the part of the conductance that varies with the applied bias.\\
In the experiment, however, we would also need to realize a polarisation exceeding $70\%$ without incorporating spin-active scattering. This can be realized if we remember that the independence of interactions was derived under the assumption that the interaction is not spin-specific. Indeed, there is one example where a spin-dependent interaction in a superconductor-hybrid experiment has been observed. If we do not operate the two quantum dots shown in Fig. \ref{fig1}(b) on resonance but apply a top-gate voltage such that we are in the Kondo regime we have two superconductor-quantum dot-ferromagnet devices. These devices have been analysed in \cite{soller4}. In this work it has been shown that the superconductor-quantum dot-ferromagnet device in the deep Kondo limit can be described by a resonant level model as far as the electronic transport is concerned. However, the splitting of the Kondo resonance due to the exchange field of the ferromagnet leads to a bias tunable spin-polarisation of the current. Consequently we can describe the system using a resonant level model as in Eq. (\ref{grij}) with a voltage-dependent polarisation of the ferromagnet that reaches $P\approx 70\%$ in the experiment \cite{soller4}.\\
Additionally, in the deep Kondo limit a collective state consisting of the electron on the quantum dot and the Kondo screening cloud by bulk states in the ferromagnet is formed so that specifics of the interface and especially spin-active scattering is strongly suppressed \cite{soller4}. This suppression allows us to neglect the aforementioned effects spin-active scattering would introduce.\\
Finally, there are new developments in nanometer sized synthetic antiferromagnets \cite{wang,2008NatPh...4...37K}. These layered materials consist of coupled ferromagnetic-non-magnetic-ferromagnetic trilayered structures which have a fixed polarized ferromagnetic bottom layer but a free ferromagnetic top layer. The top layer spin polarization can be adjusted by small local magnetic fields which permits to achieve the necessary tunability of magnetisation. In principle they could also allow for fast switching, necessary to close loopholes in Bell inequality violation experiments.\\
To conclude we have provided three steps towards the verification of entanglement in superconducting beamsplitters. The first step was to calculate the Bell parameter for these devices and to show that all necessary information can be obtained from a typical Y-junction setup. In the second step we have shown that, if the measurements are to be trusted, the degree of polarisation needed to verify entanglement in such experiments is much lower than the polarisation needed to violate a Bell inequality. These steps allowed us in the third step to show that such experiments can actually be realized by exploiting the latest technological advances in on-chip electronics and measurement techniques. This way we overcome the need for measuring higher cumulants or time-resolved measurement schemes. Such experiment would also pave the way towards on-chip quantum computation \cite{PhysRevA.57.120}.

\end{document}